\documentclass[amsmath,amsfonts,twocolumn,superscriptaddress,prx,longbibliography]{revtex4-1}
\usepackage{graphicx}
\usepackage{epsfig}
\usepackage{bm}
\usepackage{dcolumn}
\usepackage{amsmath}
\usepackage{amssymb}
\usepackage{color}

\begin{document}
\title{Fine structure of current noise spectra in nanoelectromechanical resonators}

\author{Dong E. Liu}
\affiliation{State Key Laboratory of Low Dimensional Quantum Physics, Department of Physics, Tsinghua University, Beijing, 100084, China}
\affiliation{Beijing Academy of Quantum Information Sciences, Beijing 100193, China}

\author{Alex Levchenko}
\affiliation{Department of Physics, University of Wisconsin–Madison, Madison, Wisconsin 53706, USA}

\date{September 27, 2024}

\begin{abstract}
We study the frequency-dependent noise of a suspended carbon nanotube quantum dot nanoelectromechanical resonator induced by electron-vibration coupling. Using a rigorous Keldysh diagrammatic technique, we establish a formal framework connecting the vibrational properties to electrical measurements. We find that the noise power spectrum exhibits a narrow resonant peak at the frequency of the vibrational modes. However, this fine structure tends to disappear due to a coherent cancellation effect when the tunneling barriers are tuned to a symmetric point. Notably, measuring the electrical current noise spectra provides a sensitive alternative method for detecting the damping and dephasing of quantum vibrational modes.
\end{abstract}

\maketitle

\section{Introduction} 

Nanoelectromechanical systems (NEMS) provide versatile platforms for studying quantum mechanical behaviors and hold great potential for applications in quantum sensing~\cite{AresPRL16,KhoslaPRX18,QuantumSensingRMP,ArrangoizNature2019} and quantum information processing~\cite{ClelandPRL2004,NMReadoutNature,Poot12,Palyi12,Ohm12,XiangRMP13,RipsPRL13,NoriPRL16,RossiNature2018,QinNPJQ2019,BachtoldPRX21}. The coupling effects between mechanical vibrations and the electronic degrees of freedom in a single-electron transistor have been theoretically studied in various regimes~\cite{MitraPRB04,ArmourPRB04,Blanter04,IsacssonEPL,Mozyrsky06,Usmani07,Micchi15PRL,Micchi16PRB,DEL-QE2019}. Due to their unique properties -- low mass and high stiffness -- carbon nanotubes (CNTs) are promising materials for high-quality nanomechanical applications~\cite{Sapmaz03,Sazonova04,Garcia-Sanchez07,Huttel09,Lassagne09,Steele09,HuttelPRL09,WenNP2020}. CNTs are highly sensitive to thermal and quantum fluctuating forces caused by electron hopping onto or off the resonator, which in turn induce damping, frequency shifts, and dephasing in the resonator's absorption spectrum.

At low temperatures, the amplitude of flexural vibrations can be measured by observing the DC current through the CNT. In these measurements, electrical frequency-mixing~\cite{Sazonova04,PengPRL06,Witkamp06,Lassagne09} or rectification~\cite{Huttel09,Steele09} techniques are often employed. Additionally, current noise power spectrum measurements offer highly sensitive detection~\cite{moser13,Moser2014NN}. The direct relationship between the current noise spectrum and the vibration modes provides a precise method for detecting electronic dynamics via electron-vibration coupling.

In this work, we consider a suspended CNT quantum dot (QD) resonator connected to source and drain reservoirs and study the noise power spectrum of the current flowing through the device. The electron charge density in the quantum dot is linearly coupled to the flexural vibration mode due to the vibration's influence on the gate capacitance. We show that electron-vibration coupling results in narrow peaks at the mechanical mode frequencies in the current noise power spectrum.

Using the rigorous Keldysh diagrammatic Green's function formalism for nonequilibrium systems, we reveal that these narrow peaks arise from electron-vibration vertex corrections to the disconnected diagrams (in the case without electron-vibration coupling). A similar noise peak was previously found in the classical regime by solving rate equations~\cite{armour04}. Here, we directly link the noise spectrum to the Green's function of the vibration mode in both quantum and classical regimes.

The interplay between electron dynamics and electron-vibration coupling leads to the decay and frequency shift of the narrow peaks. Thus, even in the weak coupling limit, measuring the current noise spectrum offers a sensitive method to probe the electronic degrees of freedom and to detect the associated thermal and quantum noise properties. Surprisingly, we find a robust coherent cancellation effect at a symmetric coupling point, where certain important vibrational contributions decouple from the transport processes, resulting in the disappearance of the fine structure.

\begin{figure}
  \centering
 \includegraphics[width=3.25in]{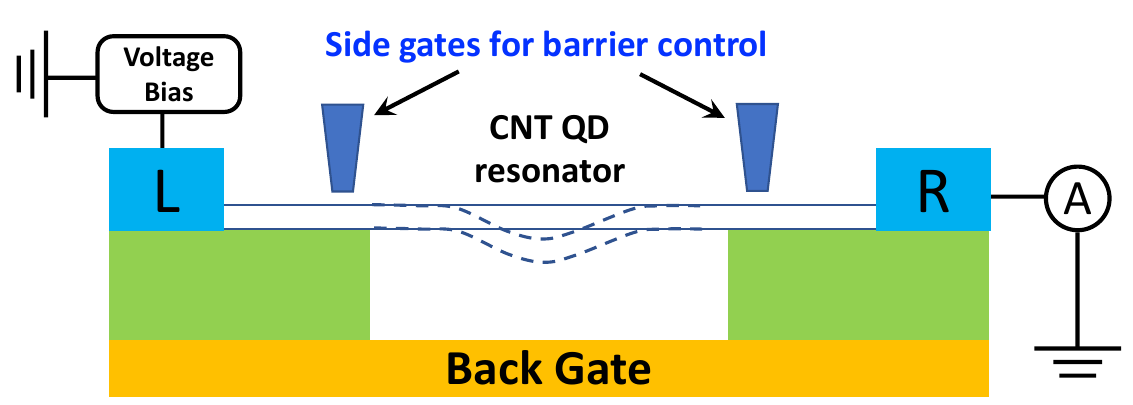}
 \caption{Schematic of the device: A suspended, doubly-clamped semiconducting CNT quantum dot (QD) is connected to the source (left) and drain (right) leads. A back gate is used to control the energy levels of the QD, and two side gates control the tunneling barriers.}\label{Fig-CNTQD}
\end{figure}

\section{Model and assumptions}

Let us consider a suspended, doubly-clamped carbon nanotube quantum dot coupled to left and right metallic leads (see Fig. \ref{Fig-CNTQD}). The quantum dot is assumed to be in the Coulomb blockade regime. The flexural vibrations of the nanotube couple to the electronic degrees of freedom through their effect on the gate capacitance. The coupling between electrons and other vibrational modes, such as breathing and stretching modes, is very weak and can therefore be neglected. The Hamiltonian of the CNT QD can be described by a constant interaction model $H_{D}=E_{c}(N-N_{g})^{2}$,
where $E_{c}=e^{2}/(2C_{\Sigma})$ is the Coulomb charging energy with the
total capacitance $C_{\Sigma}=C_{l}+C_{r}+C_{g}$, and  $N=\sum_{i\sigma}d_{i\sigma}^{\dagger}d_{i\sigma}$ is the electron number operator of the dot. Here $N_{g}=V_{g}C_{g}/e$
denotes the background charge, where $V_{g}$ is the gate voltage and 
$C_{g}$ is the gate capacitance. 

We focus on the lowest flexural vibrational mode, which can be described by a harmonic oscillator model, $H_{vib}=p^{2}/2M+M\omega_0^2 q^2 /2$, 
where $M$ is the mass of the CNT quantum dot, $\omega_{0}$ is the resonance frequency of the mode, and $q$ is the displacement of the CNT away from its equilibrium position.

In the presence of the flexural vibrations, the gate capacitance becomes
a function of the displacement coordinate $q$,
$C_{g}(q)=C_{g}^{0}+\partial_{q}C_{g}^{0} q+\partial_{q}^{2}C_{g}^{0} q^{2}/2+\cdots$
Since
$(\partial_{q}C_{g}^{0}q )/(\partial_{q}^{2}C_{g}^{0} q^{2}/2)\sim (d/q)\gg1$,
where $d$ is the distance between the CNT and the gate, 
we can neglect the $q^{2}$ term and other higher order terms. Then,
the dot Hamiltonian becomes \cite{Sapmaz03,Lassagne09,Steele09}
\begin{equation}
H_{D}=E_{c}(N-N_{g}^{0})^{2}-E_{c}\frac{2V_{g}}{e}\partial_{q}C_{g}^{0}\, qN.
\end{equation}
In what follows we consider (i) the regime $\max(T,\, V_{SD})\gg T_{K}$, here $T_{K}$ is the Kondo temperature, $V_{SD}$ is the bias voltage, 
such that the Kondo physics is irrelevant; and (ii) assume the condition $(T,\, V_{SD})\ll\Delta<E_{c}$, where $\Delta$ is the mean level spacing of the dot, so that only single energy level $\epsilon_{d}$ near the Fermi energy is relevant. 
Including the left ($\alpha=l$) and right ($\alpha=r$) leads (with the creation operator $c^\dagger_{\alpha k}$ at mode $k$) and the tunneling between
the leads and CNT QD, the Hamiltonian takes the form 
\begin{eqnarray}
H&=&\sum_{\alpha,k}\epsilon_{k}c_{\alpha k}^{\dagger}c_{\alpha k}+\epsilon_{d}d^{\dagger}d+\lambda d^{\dagger}d\, (a+a^{\dagger})\nonumber\\
  &&+\sum_{\alpha,k}\left(V_{\alpha,k}c_{\alpha k}^{\dagger}d+h.c.\right)+\omega_{0}a^{\dagger}a.
\label{eq:EV_H}
\end{eqnarray}
Here we presented the vibrational part in the second quantized form: $H_{vib}=\omega_{0}a^{\dagger}a$
, and $q=(a+a^{\dagger})/\sqrt{2M\omega_{0}}$,
thus coupling constant $\lambda=-E_{c}\frac{2V_{g}}{e}\partial_{q}C_{g}^{0}/\sqrt{2M\omega_{0}}$, while $d^\dagger$ is the fermion creation operator on the QD level. Note that throughout the paper we choose units $\hbar=k_B=1$.

\section{Current noise spectrum}

The current through the left/right junction, induced by electron tunneling events, can be expressed in the standard form using the commutator of the Hamiltonian and the particle number operator.
\begin{eqnarray}
I_{\alpha}  =  ie[H, N_{\alpha}]=ie\sum_{k}\left(V_{\alpha k}c_{\alpha k}^{\dagger}d-V_{\alpha k}^{*}d^{\dagger}c_{\alpha k}\right).
\end{eqnarray}
In the experiments, one measures the current through either the left or the right lead, which does not correspond to the naively calculated single-particle tunneling junction currents. Indeed, for finite-frequency noise measurements, the properly defined current must include additional capacitive effects and can be written as \cite{Hanke95,Blanter01}:
\begin{equation}
 I(t) = \frac{C_r+C_g}{C_l + C_r+C_g} I_{l}(t) - \frac{C_l}{C_l + C_r+C_g} I_{r}(t)
 \label{eq:leadcurrent}
\end{equation}
For $C_{l}=C_{r}+C_g$, the symmetrized current noise correlation function is defined as
\begin{eqnarray}
S(t,t') && =  \langle\left\{ \delta I(t),\delta I(t')\right\} \rangle\nonumber \\
 && =  \frac{1}{4}\Big(\langle\left\{ \delta I_{L}(t),\delta I_{L}(t')\right\} \rangle+
                \langle\left\{ \delta I_{R}(t),\delta I_{R}(t')\right\} \rangle\nonumber \\
  &&           - \langle\left\{ \delta I_{L}(t),\delta I_{R}(t')\right\} \rangle-
             \langle\left\{ \delta I_{R}(t),\delta I_{L}(t')\right\} \rangle \Big).
\label{eq:TotalNoise}
\end{eqnarray}
The noise power spectrum $S(\omega)$ can be then obtained from the Fourier transformation. Using the Keldysh Green's function formalism, see Ref. \cite{Kamenev09} for the review, the noise correlator can be written as
\begin{equation}
S_{\alpha \alpha'}(t,\, t')= \langle\left\{ \delta I_{\alpha}(t),\delta I_{\alpha'}(t')\right\} \rangle
   =C_{\alpha \alpha'}^{K}(t,t')-2\langle I_{\alpha}\rangle \langle I_{\alpha'}\rangle
\label{eq:NoiseF3}
\end{equation}
where $C_{\alpha \alpha'}^{K}$ is the Keldysh component of the contour-ordered Green's function \cite{Haug&JauhoBook}
 \begin{align}
 &C_{\alpha \alpha'}(t,t') =
 e^2\sum_{kk'}|V_{\alpha k}|^{2}|V_{\alpha' k'}|^{2}\oint_{\mathcal{C}}dt_{1}dt_{2}K_{\alpha\alpha'}(t,t',t_1,t_2)\nonumber \\
 &+e^{2}\sum_k |V_{\alpha k}|^{2}\left[g_{\alpha k}(t',t)G_{dd}(t,t')+g_{\alpha k}(t,t')G_{dd}(t',t)\right] \delta_{\alpha\alpha'}.
\label{eq:CLL}
\end{align}
In this expression, we introduced the single-particle Green's function in the leads $g_{\alpha k}$, and corresponding full Green's function on the dot  
$G_{dd}(t,t')=-i\langle T_{c} d(t) d^{\dagger}(t') \rangle$. In accordance with the standard convention $T_c$-operator means time ordering along the time contour $\mathcal{C}$. 
The four-point time correlation function under the double-integral is found in the form  
\begin{align}
K_{\alpha\alpha'}=
\Big[&-g_{\alpha k}(t_{1},t)\, g_{\alpha' k'}(t_{2},t')\,\mathfrak{D}_{1}(t,t',t_{1},t_{2})\nonumber \\ 
&+g_{\alpha k}(t_{2},t)\, g_{\alpha' k'}(t',t_{1})\,\mathfrak{D}_{2}(t,t',t_{1},t_{2})\nonumber \\
&-g_{\alpha k}(t,t_{1})\, g_{\alpha' k'}(t_{2},t')\,\mathfrak{D}_{3}(t,t',t_{1},t_{2})\nonumber \\ 
&-g_{\alpha k}(t,t_{1})\, g_{\alpha' k'}(t',t_{2})\,\mathfrak{D}_{4}(t,t',t_{1},t_{2})\Big],
\end{align}
where we introduced full two-particle Green's functions
\begin{eqnarray}
\mathfrak{D}_{1}(t,t',t_{1},t_{2}) & = & i^{2}\langle T_{c}d(t)\, d(t')\, d^{\dagger}(t_{1})\, d^{\dagger}(t_{2})\rangle,\nonumber\\
\mathfrak{D}_{2}(t,t',t_{1},t_{2})& =& i^{2}\langle T_{c}d(t)\, d^{\dagger}(t')\, d(t_{1})\, d^{\dagger}(t_{2})\rangle,\nonumber\\
\mathfrak{D}_{3}(t,t',t_{1},t_{2}) & = & i^{2}\langle T_{c}d^{\dagger}(t)\, d(t')\, d(t_{1})\, d^{\dagger}(t_{2})\rangle,\nonumber\\
\mathfrak{D}_{4}(t,t',t_{1},t_{2}) & = & i^{2}\langle T_{c}d^{\dagger}(t)\, d^{\dagger}(t')\, d(t_{1})\, d(t_{2})\rangle.\label{D-dddd}
\end{eqnarray}
Thus far, these expressions are exact with respect to electron-vibration coupling. For this reason, Wick's theorem does not straightforwardly apply, as the Green's functions are written in the explicit interaction representation. To make further progress, we need to evaluate the full two-particle Green's function. Since this is generally not possible, we must rely on a reasonable approximation scheme for controlled calculations.

\begin{figure}
\centering
\includegraphics[width=3.25in]{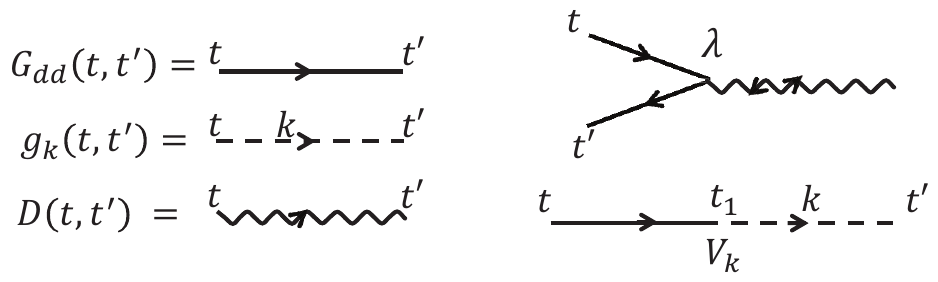}
\caption{Building blocks of diagrammatic perturbation theory. Left panel: The solid line represents the full particle Green's function $G_{dd}$ on the dot, the dashed line corresponds to the single-particle Green's function 
$g_{k}$ in the leads, and the wavy line stands for the phonon (vibration) Green's function propagator, $D(t,t')=-i\langle T_{c}(a^{\dagger}(t)+a(t))(a^{\dagger}(t')+a(t'))\rangle$. 
Right panel: Vertices for the electron-vibration coupling $\lambda$ and electron tunneling $V$.}
\label{fig:DiagrammaticR}
\end{figure}

\subsection{Narrow peaks from coupling to vibrations}

We use perturbation theory to express the noise correlation function in terms of the diagrams shown in Fig. \ref{fig:DiagrammaticR}. 
In the absence of electron-vibration interaction, the two-particle Green's functions reduce to products of single-particle Green's functions. For small electron-vibration coupling $\lambda\ll\Gamma$, 
where $\Gamma=\Gamma_L+\Gamma_R$ and $\Gamma_i=\pi\sum_{k}|V_{i k}|^2\delta(\omega-\epsilon_k)$, $\lambda/\Gamma$ can be treated as a control parameter in the perturbative expansion.

\begin{figure}
\centering
\includegraphics[width=3.25in]{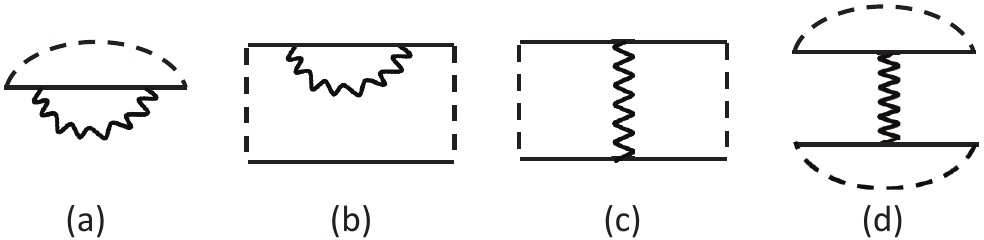}
\caption{Leading-order correction terms induced by electron-vibration coupling in diagrammatic perturbation theory.}
\label{fig:Corrections}
\end{figure}

In this approach \cite{Haug&JauhoBook}, the contribution from the disconnected part of the diagrams exactly cancels the $-2\langle I_{L}\rangle^{2}$
term in the current noise formula. The leading-order corrections to the noise function $C_{\alpha\alpha'}$, due to electron-vibration coupling, are given by four classes of connected diagrams, shown in Fig.~\ref{fig:Corrections}.

In the weak-coupling limit, corrections (a) and (b) from Fig.~\ref{fig:Corrections} simply renormalize the particle Green's function $G_{dd}$, shifting the position of the resonant energy level and broadening its width, thereby changing the lifetime of the 
$d$-electron, $\sim \lambda^2/\Gamma$. As a result, these terms do not produce significant changes in the current noise spectrum. In the opposite, strong-coupling limit $\lambda>\Gamma$, the strong electron-vibration coupling narrows the spectral peak due to the Franck-Condon blockade \cite{Koch05} and induces satellite peaks at phonon frequencies \cite{Flensberg03}.

In the following, we assume the CNT resonator has a high quality factor, and the bare phonon attenuation $\gamma_{ph}$ is much smaller than the electron tunneling rate. The additional lifetime due to electron-vibration coupling is also very small for $\lambda\ll\Gamma$. Thus, the phonon propagator in the correction term for diagram (c) behaves effectively as a $\delta$-function, reducing this correction to a convolution of the electron Green's functions 
$G_{dd}$ and $g_{\alpha k}$. Consequently, the corrections to the noise spectrum from diagram (c) produce only broad, shallow spectral peaks with a width of $\sim\Gamma$, without sharp features.

The diagram type (d) in Fig.~\ref{fig:Corrections} can be divided into two parts after cutting the vibration propagator. Therefore, processes of type (d) primarily result in sharp resonant effects, mostly due to the modified vibration mode itself.

\begin{figure}
\centering
\includegraphics[width=3.25in]{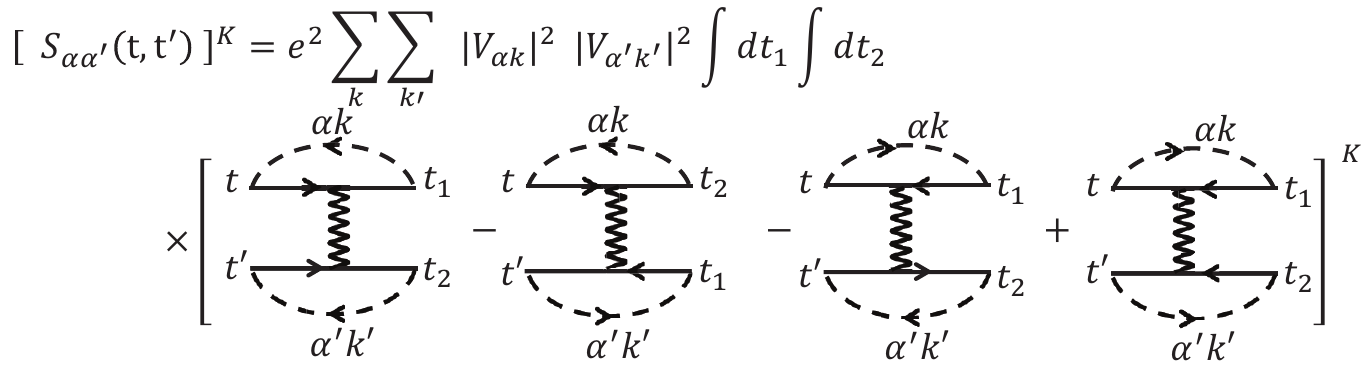}
\caption{A diagrammatic representation of the resonant correction to the noise power spectrum arising from the four-point correlation function.}
\label{fig:Corrections-d}
\end{figure}

The analytical expression for the correction to the noise spectrum from diagram (d) consists of four individual contributions, each arising from different time arrangements as defined in Eq. \eqref{D-dddd} (see Fig. \ref{fig:Corrections-d}). 
These contributions can be conveniently presented as follows:
\begin{equation}
 S_{\alpha\alpha'}(\omega) = \frac{i}{4}\lambda^2 \sum_{\eta \eta'=1,2} D^{\eta\eta'}(\omega)\times F^{\eta\eta'}_{\alpha\alpha'}(\omega),
 \label{eq:SSS}
\end{equation}
where $F(\omega)$ depends on the electronic part of the system
\begin{eqnarray}
 F^{\eta\eta'}_{\alpha\alpha'}(\omega) &=& \int \frac{d\omega_1}{2\pi} \rm{Tr}\Big[ 
               \hat{\gamma}^2 \mathbf{G}_{dd}(\omega+\omega_1) \hat{\gamma}^{\eta} \mathbf{G}_{d\alpha}(\omega_1)\nonumber\\
          && - \hat{\gamma}^2 \mathbf{G}_{\alpha d}(\omega+\omega_1) \hat{\gamma}^{\eta} \mathbf{G}_{dd}(\omega_1)\Big]\times\nonumber\\
          && \int \frac{d\omega_2}{2\pi} \rm{Tr}\Big[ 
               \hat{\gamma}^2 \mathbf{G}_{dd}(\omega_2) \hat{\gamma}^{\eta'} \mathbf{G}_{d\alpha'}(\omega_2+\omega)\nonumber\\
          && - \hat{\gamma}^2 \mathbf{G}_{\alpha' d}(\omega_2) \hat{\gamma}^{\eta'} \mathbf{G}_{dd}(\omega_2+\omega)\Big],
\end{eqnarray}
and is a smooth function of frequency compared to the phonon part $D^{\eta \eta'}(\omega)$. The gamma matrices are $\hat{\gamma}^1=\mathbb{I}_{2\times 2}$ 
and $\hat{\gamma}^2=\mathbb{\sigma}_x$ per convention in Ref.~\cite{Kamenev09}.
The phonon matrix Green's function and the $d$-electron Green's function in the rotated retarded/advanced/Keldysh ($R/A/K$) basis have the respective structures
\begin{equation}
 \mathbf{D}(\omega)= \left[D(\omega)\right]^{ij}=\left( \begin{array}{cc}
D^{K}(\omega) & D^{R}(\omega)\\ D^{A}(\omega) & 0
\end{array}\right),
\end{equation}
and 
\begin{equation}
 \mathbf{G}_{dd}(\omega)=\left( \begin{array}{cc}
G_{dd}^{R}(\omega) & G_{dd}^{K}(\omega)\\0& G_{dd}^{A}(\omega)
\end{array}\right).
\end{equation}
The Green's functions $\mathbf{G}_{d\alpha}$ and $\mathbf{G}_{\alpha d}$ have the same structure
as $\mathbf{G}_{dd}$, and their components are
\begin{eqnarray}
&& G_{d\alpha}^{K}(\omega) = \sum_{k}|V_{\alpha k}|^2\Big(G_{dd}^{R}(\omega) g_{\alpha k}^{K}(\omega) + G_{dd}^{K}(\omega) g_{\alpha k}^{A} (\omega)  \Big), \nonumber\\
&&G_{\alpha d}^{K}(\omega) = \sum_{k}|V_{\alpha k}|^2\Big( g_{\alpha k}^{R}(\omega)G_{dd}^{K}(\omega) +  g_{\alpha k}^{K}(\omega)G_{dd}^{A}(\omega)   \Big), \nonumber\\
&&G_{d\alpha}^{R/A}(\omega)= G_{\alpha d}^{R/A}(\omega) =\sum_{k}|V_{\alpha k}|^2 g_{\alpha k}^{R/A}(\omega) G_{dd}^{R/A}(\omega).
\end{eqnarray}
The single-particle Green's functions in the leads are related to the tunneling rates as follows
$\sum_{k}|V_{\alpha k}|^2 \Im g_{\alpha k}^{R/A}(\omega)=\mp \Gamma_{\alpha}$, and 
$\sum_{k}|V_{\alpha k}|^2 g_{\alpha k}^{K}(\omega)=-2i\Gamma_{\alpha}[1-2f(\omega-\mu_{\alpha})]$,
where $f(\omega-\mu_{\alpha})$ is the Fermi distribution function of the lead-$\alpha$ 
with the chemical potential $\mu_{\alpha}$. The Green's function on the QD is taken for the case without
electron-vibration coupling: $G_{dd}^{R(A)}(\omega)=1/[\omega-\epsilon_d\pm i\Gamma/2]$
and $G_{dd}^{K}(\omega)=-i \sum_{\alpha} \Gamma_{\alpha}[1-2f(\omega-\mu_{\alpha})]
/[(\omega-\epsilon_d)^2+(\Gamma/2)^2]$. 

The function $F(\omega)$ in the noise correction is related only to the electron degrees of freedom, whose spectrum changes slowly on a large energy scale $\Gamma$. We assume that the phonon decay rate $\gamma_{ph}$ is much smaller than the electron energy scale $\Gamma$. This assumption is justified for a high-quality factor $Q_0$	of the CNT resonator, where $\omega_0/Q_0\ll\Gamma$, and for small electron-vibration coupling, $\lambda^2/\Gamma\ll\Gamma$.

\begin{figure}
\centering
\includegraphics[width=3.25in]{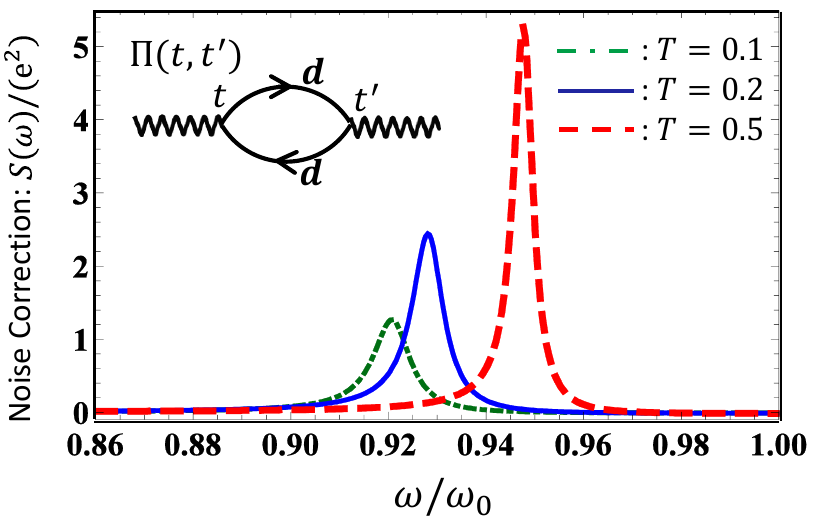}
\caption{(color online) The correction to the noise spectrum due to electron-vibration coupling is given by: $S(\omega)=(S_{LL}(\omega)+S_{RR}(\omega)-S_{LR}(\omega)-S_{RL}(\omega))/4$
for different temperatures. The parameters are $\Gamma_R=0.6$, $\mu_L=-\mu_R=0.1$, $\epsilon_d=0.0$, $\lambda=0.1$, and $\omega_0=0.05$, all chosen in units of $\Gamma_L$. 
The inset shows the diagram for the leading-order correction of the phonon polarization $\Pi$.} 
\label{fig:NoiseS_bias0_FIG}
\end{figure}

Under these conditions, the noise spectrum in Eq.~(\ref{eq:SSS}) primarily depends on the phonon propagator, which exhibits a sharp feature in the form of a narrow peak at the resonant frequency of the vibrational mode. The resonant frequency and peak width can be significantly influenced by electron-vibration coupling. We account for this effect by calculating the phonon polarization $\Pi$ up to the leading-order correction due to electron-vibration coupling, as shown in the inset of Fig. \ref{fig:NoiseS_bias0_FIG}. The phonon polarization depends on the electron Green's function and has the form \cite{MitraPRB04}:
\begin{equation}
 \Pi^{\alpha\beta}(\omega) = -\frac{i}{2}\lambda^2\int\frac{d\omega'}{2\pi} 
        \rm{Tr}\left\{ \hat{\gamma}^{\alpha}\mathbf{G}_{dd}(\omega+\omega')
         \hat{\gamma}^{\beta}\mathbf{G}_{dd}(\omega')  \right\}.
\end{equation}
The phonon Green's function is therefore
\begin{equation}
 D^{R}(\omega) = \frac{2\omega_0}{\omega^2-\omega_0^2+i\eta \rm{sgn}(\omega)-2\omega_0 \Pi^{R}(\omega)}
\end{equation}
and $D^{A}(\omega)=[D^{R}(\omega)]^{*}$. In equilibrium $\mu_L=\mu_R$, the Keldysh component is
$D^{K}(\omega)=\coth(\omega/2T)[D^{R}(\omega)-D^{A}(\omega)]$.

Figure \ref{fig:NoiseS_bias0_FIG} shows the leading-order correction to the noise spectrum, $S(\omega)=(S_{LL}(\omega)+S_{RR}(\omega)-S_{LR}(\omega)-S_{RL}(\omega))/4$, due to electron-vibration coupling. The noise spectrum exhibits a narrow peak at the resonant frequency of the CNT resonator. Strong electron tunneling at low temperatures (i.e., electrons hopping onto and off the resonator) induces a dissipative force due to the delay effect, leading to a downshift in the resonant frequency and increased damping of the resonator.

As the temperature increases, thermal fluctuations enhance the power of the current noise, as clearly seen in Fig.~\ref{fig:NoiseS_bias0_FIG}. We also conclude that the sharp peak can be used to probe the dynamics of tunneling electrons through the quantum dot resonator via electron-vibration coupling. 

\subsection{Coherent cancellation in a symmetric point}

\begin{figure}
\centering
\includegraphics[width=3.25in]{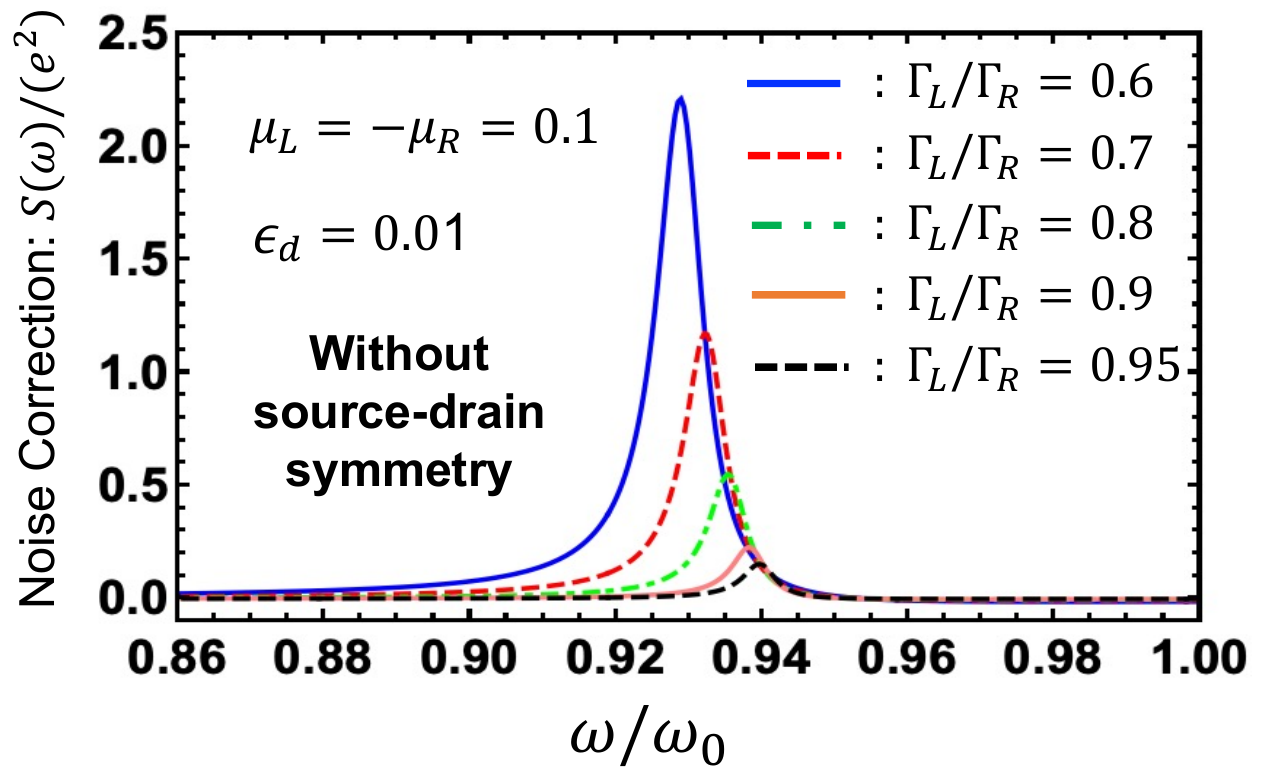}
\caption{(color online) Structure of the resonant peak 
for different junction asymmetry ratios $\Gamma_R/\Gamma_L$. Parameters: $\mu_L=-\mu_R=0.1$, $T=0.2$, $\omega_0=0.05$, $\lambda=0.1$, and $\epsilon_d=0.01$. 
All energy scales are normalized to units of $\Gamma_L$.} 
\label{fig:NoiseS_bias0_cancellationeffect_FIG}
\end{figure}

For a symmetric point where $\Gamma_L=\Gamma_R$ and $\mu_L-\epsilon_d=-\mu_R+\epsilon_d$, we find that 
$F^{\eta\eta'}=F^{\eta\eta'}_{LL}+F^{\eta\eta'}_{RR}-F^{\eta\eta'}_{LR}-F^{\eta\eta'}_{RL}=0$, leading to the disappearance of the narrow peak at the resonant frequency in the noise spectrum. 
This cancellation effect is illustrated in Fig.~\ref{fig:NoiseS_bias0_cancellationeffect_FIG}; as $\Gamma_R/\Gamma_L$ approaches unity, the narrow peak becomes weaker and eventually vanishes.
It should be emphasized that (i) this cancellation occurs only in a specific type of diagram, namely type-(d), which is responsible for the narrow peak; 
(ii) electron-vibration coupling persists even at the symmetric point, for example, through the influence on the electron Green's function.

How do we understand this cancellation effect, and does it originate from the leading-order perturbation theory? 
To answer these questions, one can apply a left-right rotation at zero voltage bias 
($\mu_L=\mu_R$) and introduce the even and odd channel operators, as described in \cite{Glazman&Pustilnik}.
\begin{equation}
\left(\begin{array}{c}
\Psi_{e}\\
\Psi_{o}
\end{array}\right)=\frac{1}{\sqrt{V_{L}^{2}+V_{R}^2}}\left(\begin{array}{cc}
V_{L} & V_{R}\\
-V_{R} & V_{L}
\end{array}\right)\left(\begin{array}{c}
\Psi_{L}\\
\Psi_{R}
\end{array}\right)
\end{equation}
where $V_{\alpha} \Psi_{\alpha}=\sum_{k}V_{\alpha k}^{*}c_{\alpha k}$ 
and $V_{\alpha}=\sqrt{\sum_k |V_{\alpha k}|^2}$.
After this rotation, the tunneling Hamiltonian becomes
\begin{align}
H_{T}=&\sum_{\alpha,k}\left(V_{\alpha,k}c_{\alpha k}^{\dagger}d+h.c.\right) \nonumber \\ 
     &=\sqrt{|V_{L}|^{2}+|V_{R}|^2}\left(\Psi_{e}^{\dagger}d+h.c.\right),
\end{align}
and the lead Hamiltonian reduces the summation of even and odd parts by the same transformation
for all momenta. Only the even channel couples to the dot and the odd channel decouples.

The current in the measurement for a general case (shown in Eq.~(\ref{eq:leadcurrent})) 
only depends on the odd channel operator
\begin{equation}
I=a I_L- b I_R=-ie\frac{V_L V_R}{\sqrt{V_L^2+V_R^2}}\left(\Psi_{o}^{\dagger}d-d^{\dagger}\Psi_{o}\right),
\end{equation}
if the condition $\Gamma_L/\Gamma_R\equiv V_{L}^2/V_{R}^{2}=b/a$ is satisfied.
To reach this condition, in principle, we only need to fine tune a single 
parameter, i.e. either $\Gamma_L$ or $\Gamma_R$, to a symmetric point $\Gamma_L/\Gamma_R\equiv V_{L}^2/V_{R}^{2}=b/a$.
When considering the noise correlation $S(t,t')=\langle\{\delta I(t),\delta I(t')\} \rangle$,
the two particle Green's functions reduce to the products of the single particle Green's functions
because the odd channel decouples. 
The noise spectrum at the symmetric point $S_{\text{SYM}}(\omega)$ becomes
\begin{align}
&S_{\text{SYM}}(\omega)= \frac{e^2}{2}\sum_{k}|V_{k}|^2\int\frac{d\omega'}{2\pi} 
     \Bigg[ g_{ok}^{<}(\omega'_-)G_{dd}^{>}(\omega'_+)+\nonumber \\
  & g_{ok}^{>}(\omega'_+)G_{dd}^{<}(\omega'_-)+g_{ok}^{<}(\omega'_+)G_{dd}^{>}(\omega'_-)
  +g_{ok}^{>}(\omega'_-)G_{dd}^{<}(\omega'_+)\Bigg]
\end{align}
where $\omega'_\pm=\omega'\pm\omega/2$ and $g_{ok}$ notation is used for the free electron in the odd channel. 
There is no two-particle Green's function contribution, such as the second and third lines of Eq.~(\ref{eq:CLL}), 
for the symmetric point. The electron-vibration coupling only modifies the electron Green's functions, with the leading-order correction 
shown in Fig.~\ref{fig:Corrections}, and cannot lead to narrow peaks at any order of perturbation theory.

For finite voltage bias, the narrow peak does not completely disappear, even at the symmetric point, but it is suppressed as the system approaches the left-right symmetric configuration, as shown in Fig.~\ref{fig:NoiseS_bias0_cancellationeffect_FIG}. 
Interestingly, this coherent cancellation effect occurs as long as the source-dot-drain voltage is in the symmetric configuration, which can be confirmed using the analytic formula for $S(\omega)$.

\section*{Acknowledgments} 

We thank Mark Dykman for discussions that helped to formulate the problem considered in this work and for the interest to this project as it evolved. D.E.L. acknowledges the support from NSF-China (Grant No.11974198). This work at the UW-Madison (A.L.) was supported by the U.S. Department of Energy (DOE), Office of Science, Basic Energy Sciences (BES) under Award No. DE-SC0020313.

\end{document}